\documentclass[twocolumn,showpacs,preprintnumbers,,superscriptaddress]{revtex4}
\usepackage{graphicx}
\usepackage{epsf}
\usepackage{dcolumn}
\usepackage{bm}
\usepackage{verbatim}
\usepackage{amssymb}
\begin{document}

\preprint{APS/123-QED}

\title{Magnetic frustration in the spinel compounds GeCo$_2$O$_4$ and GeNi$_2$O$_4$}
\author{S. Diaz}
\affiliation{Grenoble High Magnetic Field Laboratory, CNRS, BP
166, 38042 Grenoble, France}
\author{S. de Brion}
\affiliation{Grenoble High Magnetic Field Laboratory, CNRS, BP
166, 38042 Grenoble, France}
\author{G. Chouteau}
\affiliation{Grenoble High Magnetic Field Laboratory, CNRS, BP
166, 38042 Grenoble, France}
\author{B. Canals}
\affiliation{Laboratoire Louis N\'{e}el, CNRS, BP 166, 38042
Grenoble, France}
\author{V. Simonet}
\affiliation{Laboratoire Louis N\'{e}el, CNRS, BP 166, 38042
Grenoble, France}
\author{P. Strobel}
\affiliation{Laboratoire de Cristallographie, CNRS, BP 166, 38042
Grenoble, France}

\date{\today}

\begin{abstract}

In both spinel compounds GeCo$_2$O$_4$ and GeNi$_2$O$_4$ which
order antiferromagnetically (at $T_N$=23.5~K and $T_{N_1}$=
12.13~K/$T_{N_2}$=11.46~K) with different Curie Weiss temperatures
($T_{CW}$=80.5~K and -15~K), the usual magnetic frustration
criterion $f=|T_{CW}|/T_N>>1$ is not fulfilled. Using neutron
powder diffraction and magnetization measurements up to 55~T, both
compounds are found with a close magnetic ground state at low
temperature and a similar magnetic behavior (but with a different
energy scale), even though spin anisotropy and first neighbor
exchange interactions are quite different. This magnetic behavior
can be understood when considering the main four magnetic exchange
interactions. Frustration mechanisms are then enlightened.

\end{abstract}


\maketitle

Frustration is one of the most important problems of magnetism.
First proposed by G. Toulouse \cite{Toulouse} to explain the spin
glass behavior, the concept of frustration has been generalized to
systems where all the existing interactions cannot be
simultaneously satisfied leading to a highly degenerated ground
state. One of the main problem is to understand how this
degeneracy is lifted at low temperature. Many systems have been
studied \cite{Greedan} and it is clear that no general rule
exists, each system " finding its own solution ": second order
antisymetric interaction (Dzialoshinsky-Moryia) \cite{Dzialo},
ionic anisotropy \cite{ion anisotropy}, coupling to the lattice
\cite{Spin Peierls}, etc. Complex magnetic structures are often
observed i.e. with partially ordered magnetic sites.

\begin{figure}
\includegraphics[width=6.3cm]{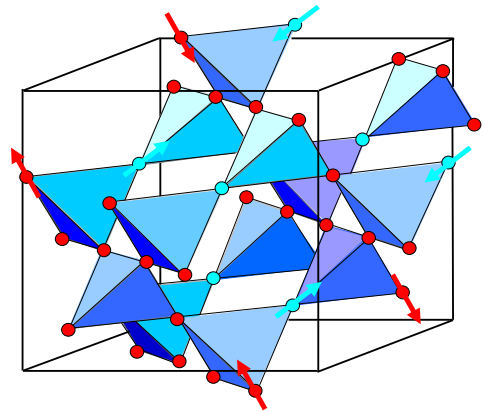}
\caption{\label{Figure 1}(Color online) Network of Magnetic ions
in the cubic spinel structure showing the corner-sharing
tetrahedra. Kagom\'e (in red) and triangular (in blue) planes are
stacked along the diagonal of the cube. The AFM structure observed
in GeNi$_2$O$_4$ is also depicted (see text for details).}
\end{figure}

There exists two types of frustration mechanisms. In the first one
frustration has its origin in the geometrical arrangement of
interactions as in triangular or tetrahedral lattices with first
neighbor antiferromagnetic (AFM) interaction; in the second one
frustration originates from competing interactions. A good example
is the square lattice with ferromagnetic (FM) interactions along
the sides of the square and a strong AFM interaction along the
diagonals. The usual criterion for frustration is the ratio
$f=|T_{CW}|/T_N>>1$ between the Curie-Weiss temperature $T_{CW}$
which is a measure of the interactions energy scale and the
N\'{e}el temperature $T_N$ below which long range order prevails.
It is generally admitted that for strongly frustrated systems $f >
10$.
\\ In this context,the  spinel compounds AB$_2$O$_4$ are
objects of renewed interest: they consist of corner-sharing
tetrahedra of magnetic ions B on a pyrochlore lattice (see Fig.
\ref{Figure 1}) and are therefore considered as good candidates
for geometrical frustration. Here we study the cases A=Ge and B=Ni
or Co. According to the usual criterion, $f > 10$, they should not
exhibit magnetic frustration. They present different spin
anisotropy and first neighbor exchange interactions. In spite of
this, we show that they have a close magnetic structure at low
temperature and a close magnetic behavior at high field. This will
be interpreted through a careful analysis of the exchange paths.
Several frustration mechanisms will be enlightened that are not
usually considered in these compounds. They arise both from
geometry and competing interactions.

Polycrystalline samples were prepared by solid state reaction.
Mixtures of GeO$_2$ with NiO or Co$_3$O$_4$ in appropriate
proportions were heated three times for 18h at 940-1000$^{\circ}$C
(Co compound) or 1140$^{\circ}$C (Ni compound) in air, with
intermittent grinding. We checked by  X-ray measurements that both
compounds adopt a normal spinel structure in the $Fd3m$ space
group with $Ge^{4+}$ ions in oxygen tetrahedral sites and magnetic
ions ($Co^{2+}$ or $Ni^{2+}$) in oxygen octahedral ones, with no
inversion. The network of magnetic ions tetrahedra is regular.

The magnetic susceptibility of both compounds present a
Curie-Weiss behavior at high enough temperature
(300~K-800~K)\cite{Diaz} indicating that $Co^{2+}$ is in a high
spin state (spin $S=3/2$ and average gyromagnetic factor $g=2.45$)
and $Ni^{2+}$ in its usual $S=1$ state with $g=2.34$. The spin
anisotropy is expected to be specific for each ion as well as the
orbital occupancy \cite{Herpin}. In these spinels, due to the
octahedral oxygen crystal field, the degeneracy of the five $d$
orbitals is lifted with three $t_{2g}$ orbitals with smaller
energy and two $e_g$ orbitals with higher energy. The orbital
occupation for the $Ni^{2+}$ and $Co^{2+}$ ions differ by a hole
in the $t_{2g}$ orbitals. This changes substantially the first
neighbor interaction as first indicated by their Curie-Weiss
temperature: it is positive for GeCo$_2$O$_4$ (+80.5~K) but
smaller and negative for GeNi$_2$O$_4$ (-15~K). The AFM long range
order occurs at $T_N=23.5$ K for the Co compound and in two steps,
at $T_{N_1}= 12.13$ K and $T_{N_2}= 11.46$ K, for the Ni compound,
as deduced from the cusp in the susceptibility, in agreement with
the literature \cite{Bertaut},\cite{Crawford},\cite{Lancaster}. We
have performed magnetization measurements up to 23~T at the
Grenoble High Magnetic Field Laboratory and in pulsed magnetic
field up to 55~T at the Laboratoire National des Champs
Magn\'etiques Puls\'es. Neutron powder diffraction patterns were
collected on the high resolution G4.1 and G4.2 diffractometers at
the Laboratoire L\'eon Brillouin, in the temperature range
$1.5-50$K.

The AFM long range order is characterized, for both compounds, by
a strong robustness against magnetic fields: $55$~T is not
sufficient to align all the magnetic moments at $4$~K (see Fig.
\ref{Figure 1}). The most striking feature in the magnetization
curves is the presence of two spin reorientation transitions at
high fields ($H_1=4.25$ T, $H_2=9.70$ T for GeCo$_2$O$_4$ and
$H_1=30$ T, $H_2=37$ T for GeNi$_2$O$_4$) revealing a close
behavior. The field scale is higher in the Ni compound ruling out
an effect of spin anisotropy (smaller for Ni) since for simple
uniaxial antiferromagnets the spin flop field is an increasing
function of the anisotropy \cite{Herpin}.

\begin{figure}
\includegraphics[width=7cm]{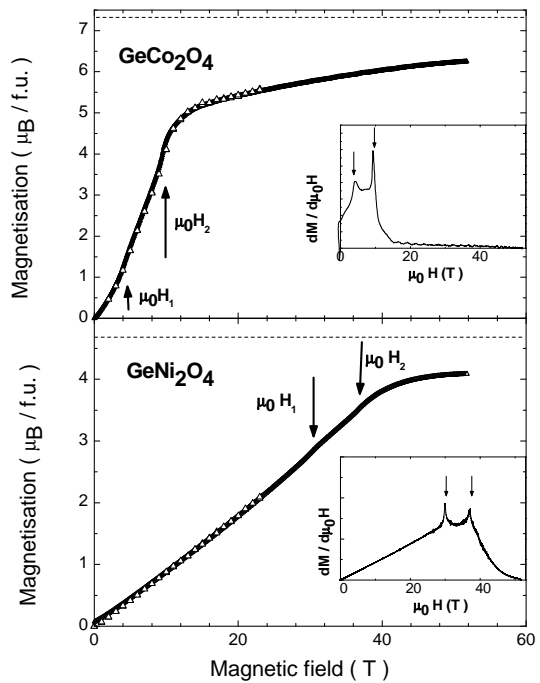}
\caption{\label{Figure 2} Magnetization versus magnetic field in
DC fields ($\triangle$ up to 23~T) and  pulsed fields
($\blacktriangle$ up to 55~T) at 4~K, for GeCo$_2$O$_4$ (a) and
GeNi$_2$O$_4$ (b). the dotted lines indicate the expected
saturation values. Inserts: derivative of the magnetization curves
enhancing the two spin flop fields. The absolute value of the
magnetization was calibrated in the DC fields experiment.}
\end{figure}

\begin{figure}
\includegraphics[width=6.9cm]{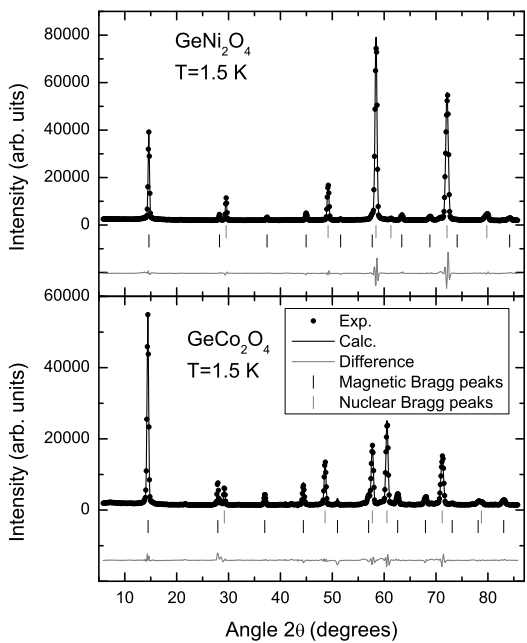}
\caption{\label{Figure 3} Rietveld refinement of the neutron
powder patterns at 1.5 K for GeCo$_2$O$_4$ and GeNi$_2$O$_4$ using
the Fullprof package \cite{fullprof} (see text).}
\end{figure}

The neutron diffraction measurements show also a close behavior
for the Ni and Co compounds (Fig. \ref{Figure 3}). The additional
Bragg peaks rising below $T_N$, characteristics of the AFM phase,
can be indexed with the same propagation vector $\vec k=
(1/2,1/2,1/2)$. There are four magnetic atoms in the asymmetric
unit chosen on the same tetrahedron. Three of them belong to a
kagom\'e plane and the fourth one to a triangular plane. These
spin orientations are in agreement with the proposition by Bertaut
{\it et al.} \cite{Bertaut} rather than the one by Plumier
\cite{Plumier}. Considering the kagom\'e and triangular planes
stacked along  the (111) direction (Fig. \ref{Figure 1}), all the
spins in one given plane are parallel to each other and
antiparallel from one kagom\'e (or triangular) plane to another.
The kagom\'e site moments are in the plane perpendicular to the
(111) direction. For the triangular site moments, two different
orientations of the moments, parallel or perpendicular to the
(111) direction, yield close solutions. The diffraction pattern is
slightly better fitted with the moments parallel (GeNi$_2$O$_4$)
or perpendicular (GeCo$_2$O$_4$) to the (111) direction. The exact
orientation in this plane cannot be determined from powder
diffraction. The refined moments are reported in Table \ref{Table
1}. For GeCo$_2$O$_4$, the amplitude of the magnetic moments is
found identical for the four magnetic sites but is substantially
smaller than the value deduced from the high temperature
susceptibility and high field magnetization: $3.01~\mu_B$ instead
of $3.67~\mu_B$. For GeNi$_2$O$_4$, it is quite close for the
kagom\'e site moments: $2.29~\mu_B$ instead of $2.35~\mu_B$ but
reduced for the triangular site moment, $1.10~\mu_B$. These
reduced magnetic moments are reminiscent of what is observed in
the other compound with the same magnetic network GdTi$_2$O$_7$,
where one spin out of four on average is not ordered \cite{GdTi2O7
Neutron1}. There, the magnetic structure is described by the same
propagation vector but the alignment of the spins in one given
plane adopts the $120^{\circ}$ structure typical of triangle-based
networks of frustrated antiferromagnets \cite{Petrenko}. Note that
the presence of two decoupled AFM sets of FM sublattices observed
in the Co and Ni spinels is in agreement with the presence of two
reorientation fields in the magnetization process, one
corresponding to each set of sublattices.

\begin{table}
\caption{\label{Table 1} Amplitude of the magnetic moments, in
GeNi$_2$O$_4$ and GeCo$_2$O$_4$ with lattice parameter $a=8.340$
\AA~ and $8.251$ \AA~ respectively, determined from Rietveld
refinement of the neutron diffraction powder patterns at 1.5~K.}
\begin{ruledtabular}
\begin{tabular}{cccccccc}
$GeCo_2O_4$&\textit{x}&\textit{y}&\textit{z}&$\mu$($\mu_B$)\\
\hline
 $Co_1$& 0.00 & 0.25 & 0.75 & 3.02(1)\\
 $Co_2$& 0.25 & 0.25 & 0.50 & 3.02(1)\\
 $Co_3$& 0.25 & 0.00 & 0.75 & 3.02(1)\\
 $Co_4$& 0.00 & 0.00 & 0.50 & 3.02(1)\\
 \multicolumn{5}{l}{Nuclear phase: Bragg R factor = 2.05, Rf factor = 1.07}\\
 \multicolumn{5}{l}{Magnetic phase: R factor = 5.91}\\
 \hline
 $GeNi_2O_4$&\textit{x}&\textit{y}&\textit{z}&$\mu$($\mu_B$)\\
\hline
 $Ni_1$& 0.00 & 0.25 & 0.75 & 2.29(2)\\
 $Ni_2$& 0.25 & 0.25 & 0.50 & 2.29(2)\\
 $Ni_3$& 0.25 & 0.00 & 0.75 & 2.29(2)\\
 $Ni_4$& 0.00 & 0.00 & 0.50 & 1.10(20)\\
  \multicolumn{5}{l}{Nuclear phase: Bragg R factor = 2.59, Rf factor = 1.78}\\
  \multicolumn{5}{l}{Magnetic phase: R factor = 8.33}\\
\end{tabular}
\end{ruledtabular}
\end{table}

\begin{figure}
\includegraphics[width=6.5cm]{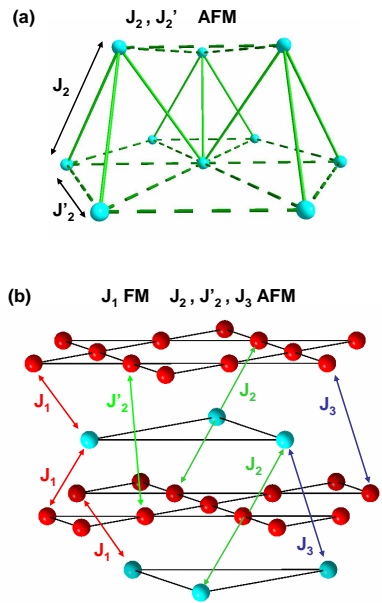}
\caption{\label{Figure 4}(Color online) Magnetic ions in the
kagom\'e (K) and triangular planes (T). (a) Geometric frustration
in the $J_2$/$J_2'$ network when only the triangular planes are
considered. The (non) satisfied ($J_2$) $J_2'$ are depicted by
(dis) continuous lines. (b) Competition between $J_1$ (whatever
its sign) and AFM $J_2$, $J_2'$ and $J_3$.}
\end{figure}

\begin{figure}
\includegraphics[width=5.8cm]{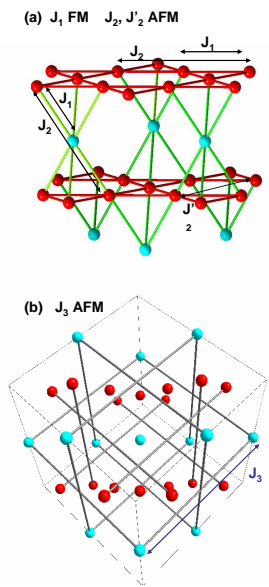}
\caption{\label{Figure 5}(Color online) (a) Competing FM $J_1$ and
AFM $J_2'$ interactions. In the (K) planes, $J_1$ dominates, in
between the (K) planes, $J_2'$ dominates. The AFM $J_2$
interaction is also depicted. (b) Network of the AFM $J_3$
interactions.}
\end{figure}

The great variety of magnetic ground states observed in
pyrochlores and spinels arises from strong magnetic frustration
mechanisms. The degeneracy of the magnetic ground state is lifted
by additional mechanisms that may differ substantially from one
compound to the other. A good knowledge of the magnetic
interactions involved in these systems is crucial for the
understanding of their magnetic properties. It implies identifying
and sorting out the different relevant magnetic exchange paths
within the magnetic network and via the oxygen network. All these
interactions depend on the magnetic and oxygen ions orbitals
through their particular occupancy and spatial distribution. We
have carefully examined the case of GeCo$_2$O$_4$ and
GeNi$_2$O$_4$ (Fig. \ref{Figure 4} and \ref{Figure 5}) using
Goodenough-Anderson-Kanamori- rules \cite{AKG rules}. The dominant
interactions between the magnetic B ions to be considered are
those from the direct exchange, the 90$^{\rm{o}}$ super exchange,
and the 135$^{\rm{o}}$ and 180$^{\rm{o}}$ super super exchange.
The direct B-B interaction is strong and FM for $Co^{2+}$ and null
for $Ni^{2+}$, respectively, due to the presence or not of a hole
in the $t_{2g}$ orbital. The 90$^{\rm{o}}$ B-O-B super exchange is
FM and weak. Both of them will be described by the effective first
neighbor interaction $J_1$. There are two different super super
exchange paths B-O-O-B at 135$^{\rm{o}}$ for the third neighbors
interaction yielding AFM J$_2'$ and J$_2$, with and without an
interstitial magnetic ion respectively. They therefore differ, in
the Co compound only, from an additional small FM direct exchange
contribution. Finally, there is the AFM 180$^{\rm{o}}$ B-O-O-B
J$_3$ interaction between the sixth neighbors, whose contribution
was also considered in Ref. \onlinecite{Bertaut}. The other
exchange paths yield negligible contributions. Note that the above
analysis rules out any geometrical frustration mechanisms within
the nearest neighbor tetrahedra since $J_1$ is FM, which is in
sharp contrast to the Gd \cite{GdTi2O7 Neutron1} or Cr compounds
\cite{Spin Peierls}.

In order to test whether these exchange path considerations alone
may account for the observed magnetic structures, we have
performed a zero temperature analysis of the wave vectors of
possible instabilities for a Heisenberg model $H = -\sum_{ij}
J_{ij} S_i \cdot S_j$ within the ($J_2/J_1, J_2'/J_1,J_3/J_1$)
phase space \cite{q-selection}.
We found that, for -2<$J_i/J_1$<+2, there are two finite regions
of parameters for which the instability wave vector belongs to the
q-star ($1/2, 1/2, 1/2$), as observed experimentally, with a
corresponding non degenerate eigenmode in reciprocal space
($1/\sqrt{3},1/\sqrt{3},1/\sqrt{3},0$) and its cyclic
permutations.
Within this approach, the ground state is described by FM
kagom\'{e} planes, as observed experimentally, and non magnetic
triangular planes. This latter feature, with one disordered site
over four in average, seems to apply partially to the Ni compound
where a reduced moment is observed for the fourth spin. Clearly,
an additional mechanism operates to select a particular
orientation for this spin, which differs in the Co and Ni spinels.
This is revealed in the orientation of the ordered part of the
fourth spin (perpendicular or parallel to (111) respectively in
both compounds); it may also be related to the two main
spin-reorientation processes in the magnetization curve which
occur at different fields and magnetization values.

In the light of the above analysis, it appears that several
frustration mechanisms are present. For instance, the network
formed by the AFM $J_2'$ and $J_2$ interactions within and in
between the triangular planes consists of tetrahedra connected by
their edges where geometrical frustration takes place (Fig.
\ref{Figure 4}(a)). Even more striking is the competition between
$J_1$ on one side (whatever its sign) which couples a kagom\'e
plane with a triangular plane and vice versa, and the AFM
interactions $J_2$, $J_2'$ and $J_3$ on the other side, which
couple identical planes (either kagom\'e or triangular) (Fig.
\ref{Figure 4}(b)). These two frustration mechanisms are relevant
for any B site magnetic spinel. For B=(Ni,Co), there is
additionally a competition between the FM $J_1$ and the AFM $J_2$
(Fig. \ref{Figure 5}(a)). Finally, the only magnetic interaction
which is always satisfied within the observed magnetic structure
is the 180$^{\rm{o}}$ super super exchange interaction $J_3$ (Fig.
\ref{Figure 5} (b)).

In conclusion, we have shown that GeCo$_2$O$_4$ and GeNi$_2$O$_4$
present an original magnetic ground state, with a (1/2,1/2,1/2)
propagation vector and four magnetic sublattices (AFM stacking of
FM Kagom\'e planes intercalated with AFM stacking of FM triangular
planes). Due to the presence of these sublattices, the field
induced common magnetic behavior of both compounds present two
spin reorientation transitions in high fields. Frustration
mechanisms take place in these systems although not as
straightforward as in other spinels and pyrochlores with AFM first
neighbor interactions. This study underlines the need to consider
exchange paths for the magnetic interactions that are well beyond
first neighbors in systems dominated by magnetic frustration.

The GHMFL is associated to the Universit\'{e} Joseph
Fourier-Grenoble I. We thank H. Rakoto (LNCP) and J.
Rodriguez-Carvajal (LLB) for the magnetisation and neutron
diffraction measurements.

\end{document}